\documentstyle [aps,psfig]{revtex}

\begin{document}
\draft
\title{Influence of scalar fields on the approach to a cosmological
singularity} 
\author{Beverly K. Berger}
\address{Department of Physics, Oakland University, Rochester, MI 48309
USA}

\maketitle
\bigskip
\begin{abstract}
The method of consistent potentials is used to explain how a minimally
coupled
(classical) scalar field can suppress Mixmaster oscillations in the
approach to the
singularity of generic cosmological spacetimes.
\end{abstract}

In their long-term study of the approach to the singularity in generic
cosmological
spacetimes
\cite{belinskii69a,belinskii69b,belinskii71a,belinskii71b,belinskii82},
Belinskii, Khalatnikov, and Lifshitz (BKL)
concluded that generic cosmological spacetimes approach the
singularity as a (different) Mixmaster universe \cite{misner69} at every
spatial point.
The Mixmaster universe is characterized by ``oscillations'' from one
Kasner solution
\cite{kasner25} to another. When BKL considered the influence of a
classical
minimally coupled scalar field on the approach to the singularity, they
found that it can
suppress Mixmaster oscillations \cite{belinskii73}. While this result is
now well-known,
the mechanism by which it happens is not widely appreciated. I wish to
address this
issue here.

The role of the scalar field may be easily clarified using the method of
consistent
potentials (MCP) originally due to Grubi\u{s}i\'{c} and Moncrief
\cite{grubisic93}. The
MCP has been applied to a variety of cosmological spacetimes to explain
the nature of the
(numerically observed) approach to the singularity in spatially
inhomogeneous cosmologies
\cite{berger98d}. The MCP first assumes that the approach to the
singularity is
asymptotically velocity term dominated (AVTD) \cite{isenberg90} and then
looks for a
contradiction. If the model is AVTD, it approaches arbitrarily closely
to a Kasner
solution with a possibly different Kasner solution at every spatial
point. For any model,
an asymptotic  velocity term dominated (VTD) solution may be found by
neglecting all
terms in Einstein's equations containing spatial derivatives and taking
the limit as
$\tau \to \infty$. (The MCP includes the implicit assumption that strong
cosmic censorship
holds for these cosmological models --- i.e.~there exists a foliation
labeled by
$\tau$ such that some curvature invariant blows up as $\tau \to
\infty$.) If the model is
actually AVTD, then substitution of the VTD solution into the full
Einstein equations
will be consistent --- asymptotically, all terms neglected in obtaining
the VTD solution
will be exponentially small. For convenience, the MCP will be applied to
the Hamiltonian
whose variation yields the relevant Einstein equations since
exponentially small
(growing) terms in the Hamiltonian will yield exponentially small
(growing) terms in
Einstein's equations upon variation. Any terms which cannot be made
consistent with the
VTD solution indicate, within the MCP, that the model is not AVTD. Here
I shall consider
essentially the same models as in \cite{berger98d} but with the addition
of a scalar
field. I shall explore the influence of a massless, minimally coupled
scalar field in
some detail. The possibly interesting dynamics which may result from
exponentially
coupled scalar fields will be discussed elsewhere. For each cosmology, I
shall use the
MCP to show how the scalar field yields an asymptotically velocity term
dominated (AVTD)
approach to the singularity.

Consider first the primary Mixmaster model --- the spatially homogeneous
vacuum Bianchi
Type IX cosmology \cite{belinskii71b,misner69}. In the presence of a
spatially homogeneous
scalar field, $\phi(\tau)$, Einstein's equations are obtained from the
variation of the
Hamiltonian constraint ${\cal H}=0$ where
\begin{equation}
\label{hix}
2{\cal H} = - p_\Omega^2 + p_+^2 + p_-^2 + p_\phi^2 + V(\Omega,
\beta_\pm) +
e^{6\Omega} {\cal V}(\phi) \ .
\end{equation}
In Eq.~(\ref{hix}), $\Omega$ is $1/3$ the logarithm of the spatial
volume and $\beta_\pm$
the anisotropic shears with $p_\Omega$ and $p_\pm$ their conjugate
momenta. The scalar
field momentum, $p_\phi$, plays a crucial role in the approach to the
singularity. The
spatial scalar curvature appears in $V(\Omega, \beta_\pm) =
{}^3\/g\,{}^3\/R$ for ${}^3\/
g$ and ${}^3\/ R$ respectively the determinant and scalar curvature of
the spatial metric
while ${\cal V}(\phi)$ is the scalar field potential. The key to the
influence of the
scalar field is the additional kinetic term in Eq.~(\ref{hix}). Any
scalar field
coupling which produces such a term will yield the same effect as long
as ${\cal
V}(\phi)$ does not contain terms exponential in $\phi$. Let us consider
\begin{equation}
\label{hixsum}
{\cal H} = {\cal H}_K + {\cal H}_V
\end{equation}
where ${\cal H}_K$ contains all the momenta and 
\begin{eqnarray}
\label{hixv}
2{\cal H}_V =&\ & e^{4\Omega-8\beta_+} + e^{4 \Omega + 4 \beta_+ + 4
\sqrt{3} \beta_-}+
e^{4
\Omega + 4 \beta_+ - 4 \sqrt{3} \beta_-} \nonumber \\
& &-2( e^{4\Omega+4\beta_+}+ e^{4 \Omega
-2\beta_+ - 2 \sqrt{3} \beta_-}+ e^{4
\Omega -2 \beta_+ +2 \sqrt{3} \beta_-})+e^{6\Omega}{\cal V}(\phi).
\end{eqnarray}
First, we note that, in these variables, the (strong curvature)
singularity in
these models occurs as $\Omega \to -\infty$. Thus, unless ${\cal
V}(\phi)$ contains
terms which asymptotically depend exponentially on $\Omega$ (through
asymptotic linear
dependence of $\phi$ on $\Omega$),
$e^{6\Omega} {\cal V}(\phi)
\to 0$ as
$\Omega \to -\infty$. For now, we shall ignore this term.

The MCP requires us to assume that ${\cal H} = {\cal H}_K$. Variation of
this Hamiltonian
yields equations with the solution
\begin{equation}
\label{bpmix}
\beta_\pm = \beta_\pm^0 + v_\pm |\Omega|,
\end{equation}
\begin{equation}
\label{phiix}
\phi = \phi^0 + v_\phi |\Omega|
\end{equation}
where $v_\pm = p_\pm/|p_\Omega|$ and $v_\phi = p_\phi /| p_\Omega|$ with
the
momenta all constant. The Hamiltonian constraint ${\cal H}
= 0$ becomes ${\cal H}_K = 0$ or
\begin{equation}
\label{hkeq0}
v_+^2 + v_-^2 + v_\phi^2 = 1.
\end{equation}
First consider the vacuum case --- $\phi \equiv 0$. Then we can write
$v_\pm$ in polar
coordinates (with unit radius) in the anisotropy plane. The
minisuperspace (MSS)
potential $V$ is dominated by the first three terms on the rhs of
Eq.~(\ref{hixv}) so that
\begin{equation}
\label{hixvapprox}
2 V \approx e^{4\Omega-8\beta_+} + e^{4 \Omega + 4 \beta_+ + 4 \sqrt{3}
\beta_-}+
e^{4\Omega + 4 \beta_+ - 4 \sqrt{3} \beta_-}.
\end{equation}
Substitution of (\ref{bpmix}), with $v_+ = \cos \theta$, $v_- = \sin
\theta$, into
(\ref{hixvapprox}) yields
\begin{equation}
\label{vixmcp}
2 V \approx e^{-4 |\Omega| (1 + 2 \cos \theta)} + e^{-4 |\Omega| (1
-\cos \theta -
\sqrt{3} \sin \theta)}+ e^{-4 |\Omega| (1 -\cos \theta +
\sqrt{3} \sin \theta)} .
\end{equation}
Except for (the set of measure zero) $\theta = \{0, 2\pi/3, 4\pi/3 \}$,
any (generic)
value of $\theta$ will cause one of the terms on the rhs of
(\ref{vixmcp}) to grow. For
example,
$1 + 2 \cos \theta < 0$ will cause the first term to grow. This
condition arises for
$\cos \theta < -1/2$ or $-2\pi/3 < \theta < 2\pi/3$.

With the addition of the scalar field, $v_\phi^2 > 0$, so that 
\begin{equation}
\label{vconditionix}
v_+^2 + v_-^2 = 1 - v_\phi^2 < 1.
\end{equation}
No term in Eq.~(\ref{vixmcp}) will grow if we can satisfy simulatneously
with
Eq.~(\ref{vconditionix})
\begin{eqnarray}
\label{vrestrix}
1 + 2 v_+ & > & 0, \nonumber \\
1 - v_+ - \sqrt{3} v_- & > & 0, \nonumber \\
1 - v_+ + \sqrt{3} v_- & > & 0.
\end{eqnarray}
Since $v_+^2 + v_-^2 = 1$ is no longer required, solution of
Eqs.~(\ref{vrestrix}) is
possible if $v_+^2 < 1/2$ and $v_-^2 < 1/12$ which can occur if $2/3<
v_\phi^2 < 1$.
Since $p_\Omega > 0$ decreases at each bounce \cite{berger96c}, any
initial value of
$p_\phi$ will eventually yield $v_\phi^2 > 2/3$. 

In the absence of the scalar field, the growing term in
Eq.~(\ref{vixmcp}) causes a
``bounce'' (see \cite{berger98d}) which changes the values of $v_\pm$
according to the
prescription first given by BKL \cite{belinskii71b}. If the new values
of $v_\pm$ satisfy
(\ref{vrestrix}), there will be no further bounces so that the solution
will approach
(\ref{bpmix})-(\ref{phiix}). We see, then, that the main function of the
scalar field is
to weaken the restriction imposed by the Hamiltonian constraint on the
gravitational
``kinetic energy.'' In Fig.~1, trajectories with and without the scalar
field are shown in
the $v_+$-$v_-$ plane. In the vacuum case, away from bounces, the
trajectory must fall on
the ``Kasner circle'' defined by $v_+^2 + v_-^2 = 1$. For a small
initial value of
$p_\phi$, the non-vacuum model closely tracks the other until the
amplitude of
$v_\phi$ grows sufficiently large. At this point, the trajectory
deviates noticeably from
the Kasner circle. Shortly thereafter, $v_\pm$ are able to satisfy
(\ref{vrestrix}) so
there are no more bounces and the values of $v_\pm$ no longer change.
(This causes the
trajectory in the $v_\pm$-plane to end.)

It is possible to retain the Mixmaster (oscillatory) behavior if ${\cal
V}(\phi)$ in
(\ref{hix}) is such that
\begin{equation}
\label{vscalarexp}
e^{6\Omega} {\cal V}(\phi) = a^2 e^{\alpha \phi} + b^2 e^{-\zeta \phi}
\end{equation}
where $\alpha, \ \zeta > 0$. A potential of this type will, according to
the MCP, yield
an exponentially growing term in (\ref{hixv}) unless $v_\phi = 0$. But
$v_\phi = 0$
means that the usual Mixmaster oscillations will occur. Thus, given
(\ref{vscalarexp}),
there will be no way to avoid oscillations. Coupling between scalar
field and
gravitational degrees of freedom in $e^{6\Omega}\, {\cal V}(\phi)$ could
lead to very
complicated behavior. BKL, in fact, by coupling an electromagnetic field
to a
Brans-Dicke-like scalar field so as to produce a potential term like
(\ref{vscalarexp})
where $a$ and $b$ are functions of the electromagnetic field, claimed to
have restored
the oscillations suppressed initially by the scalar field alone
\cite{belinskii73}. Potentials exponential in the scalar field can also
arise in string
theory \cite{lu96}.

The most complicated models to which the MCP has been applied are vacuum
cosmologies with
a single spatial $U(1)$ symmetry and $T^3$ spatial topology
\cite{moncrief86,berger93,berger98b,berger98c,berger98d}. The degrees of
freedom are
$\{x,z,\Lambda,\varphi,\omega\}$ with conjugate momenta
$\{p_x,p_z,p_\Lambda,p,r\}$.
The variables are functions of spatial coordinates $u$ and $v$ and time
$\tau$. Here the
Hamiltonian constraint (${\cal H} = 0$) is
\begin{eqnarray}
\label{hu1}
{\cal H} &=&  {1 \over 8}p_z^2+{1 \over 2}
e^{4z}p_x^2+{1 \over 8}p^2+{1 \over 2}e^{4\varphi }r^2-{1 \over
2}p_\Lambda
^2 \nonumber \\
&& +   \left( {e^\Lambda e^{ab}} \right) ,_{ab}- \left( {e^\Lambda
e^{ab}}
\right) ,_a\Lambda ,_b+e^\Lambda   \left[  \left( {e^{-2z}}
\right) ,_u x,_v- \left( {e^{-2z}} \right) ,_v x,_u \right] \nonumber
\\
&&  +2e^\Lambda e^{ab}\varphi ,_a\varphi ,_b+{1 \over 2}
e^\Lambda e^{-4\varphi }e^{ab}\omega ,_a\omega ,_b  \nonumber \\
&=&{\cal H}_K +\,{\cal H}_V 
\end{eqnarray}
where ${\cal H}_K$ contains only momenta and
\begin{equation}
\label{eab}
e^{ab}={1 \over 2}e^{-2z}\left(
{\matrix{{e^{4z}+(1-x)^2}&{-e^{4z}+(1-x^2)}\cr \cr
{-e^{4z}+(1-x^2)}&{e^{4z}+(1+x)^2}\cr
}} \right)\,.
\end{equation}
Its variation yields Einstein's equations. A transformation $\Lambda \to
\Lambda
-2\tau$ will restore the explicit time
dependence in (\ref{hu1}) found in \cite{berger93,berger98b,berger98c}.
To apply
the MCP, the appropriate generalization of (\ref{bpmix}) is required.
This is found by
solving the velocity term dominated (VTD) equations obtained by
neglecting all terms in
Einstein's equations containing spatial derivatives. In the limit as
$\tau \to
\infty$, we find
\cite{berger98b}
\begin{eqnarray}
\label{u1avtd} 
z&=&-v_z\tau,  \quad x= x_0,  \quad p_z= -4v_z, \quad
p_x= p_x^0, \quad
\varphi= -v_\varphi\tau, \nonumber  \\
\omega&=& \omega_0, \quad
p=-4v_\varphi,  \quad
r= r^0, \quad
\Lambda = \Lambda_0  - v_\Lambda\tau, \quad
p_\Lambda = v_\Lambda
\end{eqnarray}
where $v_z$, $v_\varphi$, $x_0$, $p_x^0$, $\omega_0$, $r^0$,
$\Lambda_0$, and
$v_\Lambda > 0$ are functions of $u$ and $v$ but independent of $\tau$.
(The sign of  
$v_\Lambda$ is fixed to ensure collapse.) In the $VTD$ limit, the
Hamiltonian constraint
(\ref{hu1}) becomes 
\begin{equation}
\label{hkvtdu1}
 - {1 \over 2} p_\Lambda^2 + {1 \over 8} p^2 + {1 \over 8} p_z^2 = 0.
\end{equation}
To obtain this expression, it was necessary to assume $v_\varphi > 0$
and $v_z > 0$ so
that the exponential terms containing $p_x^2\,e^{4z}$ and
$r^2\,e^{4\varphi}$  in ${\cal
H}_K$ would be exponentially small. However,
$v_\varphi > 0$ causes the term ${1 \over 2} e^{\Lambda - 4 \varphi}
e^{ab}\nabla_a \omega
\nabla_b\omega$ in (\ref{hu1}) to grow unless
\begin{equation}
\label{vcondition}
\mathop {\lim }\limits_{\tau \to \infty }\,
(\Lambda - 2 z -4 \varphi),_\tau = -v_\Lambda + 2 v_z + 4 v_\varphi < 0.
\end{equation} 
This condition is obtained by noting that $v_z > 0$ implies $z \to
-\infty$ as
$\tau \to \infty$ so that $e^{ab}$ from (\ref{eab}) behaves as
$e^{-2z}$.  As was shown in
\cite{berger98b,berger98c}, the VTD form of the Hamiltonian constraint
(\ref{hkvtdu1})
gives $v_\Lambda^2=4 v_z^2 + 4 v_\varphi^2$ so that (\ref{vcondition})
cannot be satisfied
with $v_\varphi > 0$ and
$v_z > 0$. On the other hand, if either of these is $< 0$, either $r^2
e^{4\varphi}$ or
$p_x^2 e^{4z}$ in (\ref{hu1}) will grow. This leads to the prediciton
(observed
numerically) that the approach to the singularity is oscillatory. 

Just as in the spatially homogeneous case, in the presence of a scalar
field
$\phi(u,v,\tau)$, the AVTD limit of the Hamiltonian constraint becomes
\begin{equation}
\label{hkvtdu1phi}
0 = - {1 \over 2} p_\Lambda^2 + {1 \over 8} p^2 + {1 \over 8} p_z^2 + {1
\over 2} p_\phi^2
\end{equation}
where $p_\phi$ is the scalar field momentum. It is now possible to
simultaneously satisfy
$v_\varphi > 0$, $v_z > 0$, and $-v_\Lambda + 2 v_z + 4 v_\varphi < 0$
since
$v_\Lambda^2 = 4 v_z^2 + 4 v_\varphi^2 + p_\phi^2$ from
(\ref{hkvtdu1phi})
can be made arbitrarily large with a sufficiently strong scalar field.
At any
representative spatial point, after some number of oscillations, the
momenta will move
into the range needed to make the remaining evolution AVTD at that
point. Presumably, as
$\tau \to \infty$, the model will become AVTD almost everywhere. 

It is possible to display this effect in a computer simulation of the
full Einstein
equations for $U(1)$ symmetric cosmologies with and without a scalar
field. For
convenience, $p_\phi$ is chosen to be constant in space. In Fig.~2, the
evolution of
$\varphi$ toward the singularity at three representative spatial points
is shown for both
cases. It is clear that the scalar field suppresses the oscillations. In
Fig.~3,
$\varphi(u=v,\tau)$ is shown for both cases. Note that the formation of
ever smaller
scale spatial structure is suppressed by the scalar field. Although
these particular
simulations can only be followed to $\tau \approx 40$, the MCP predicts
that, for the
scalar field model, at the spatial points of Figs.~2a and 2b, since
$v_\varphi = 0$ (to
machine accuracy) and $v_\varphi > 0$ respectively, no further bounces
will occur. At the
spatial point of Fig.~2c, at least one more bounce to change the sign of
$v_\varphi$ is
expected.

In models with two commuting spatial Killing fields, the Hamiltonian
constraint enters in
an interesting way. Consider the magnetic Gowdy model discussed in
\cite{weaver98,weaver99}. Its variables $\{P, Q, \lambda\}$ and
conjugate momenta $\{
\pi_P, \pi_Q, \pi_\lambda \}$ depend on spatial coordinate $\theta$ and
time $\tau$. 
Here, Einstein's equations are obtained by variation of a Hamiltonian
$H$ which is {\it not} the Hamiltonian constraint ${\cal H}$.
\begin{equation}
\label{hmaggowdy}
H = {1 \over {4 \pi_\lambda}} \left[ \pi_P^2 + e^{-2P}\pi_Q^2 +
e^{-2\tau} \left(
P,_\theta^2 + e^{2P} Q,_\theta^2 \right) \right] +  B^2 \pi_\lambda
e^{(\lambda + \tau)/2}
\end{equation}
where $B$ measures the strength of the magnetic field. The variation of
(\ref{hmaggowdy})
with respect to $\pi_\lambda$ yields 
\begin{equation}
\label{lamdot}
\lambda,_\tau =- {1 \over {4 \pi_\lambda^2}} \left[ \pi_P^2 +
e^{-2P}\pi_Q^2 + e^{-2\tau}
\left( P,_\theta^2 + e^{2P} Q,_\theta^2 \right) \right] +  B^2
e^{(\lambda +\tau)/2}
\end{equation}
which happens to be a rewriting of the Hamiltonian constraint ${\cal H}
= 0$. The AVTD
solution obtained from the variation of (\ref{hmaggowdy}) neglecting
spatial derivatives
is
\begin{equation}
\label{gowdyavtd}
P = v \, \tau, \quad \quad Q = Q_0, \quad \quad \lambda = - v^2 \, \tau
\end{equation}
where $v = \mathop {\lim }\limits_{\tau \to \infty } \, \pi_P/(2
\pi_\lambda)$ depends on
$\theta$ but not on $\tau$. 
As was described in \cite{berger98d}, substitution of (\ref{gowdyavtd})
into
(\ref{hmaggowdy}) shows that
$V_1 =e^{-2P} \pi_Q^2/\pi_\lambda$ will grow if $v < 0$,
$V_2=e^{2(P-\tau)}\,Q,_\theta^2/\pi_\lambda$ will grow if $v > 1$ while
$V_3 = B^2
e^{(\lambda +\tau)/2}$ will grow if $0 < v < 1$. Thus there is no value
of $v$ consistent
with the AVTD solution (\ref{gowdyavtd}).

Now consider a scalar field with momentum $\pi_\phi$. Eq.~(\ref{lamdot})
for
$\lambda,_\tau$ gains an additional term and becomes
\begin{equation}
\label{lamdotsf}
\lambda,_\tau = - {1 \over 4} \left( {{\pi_P} \over {\pi_\lambda}}
\right)^2- {1 \over 4}
\,e^{-2P}\, \left( {{\pi_Q} \over {\pi_\lambda}} \right)^2- {1 \over 4}
\left( {{\pi_\phi}
\over {\pi_\lambda}} \right)^2- {{e^{-2\tau}} \over {4 \pi_\lambda^2}}
\left(  
 P,_\theta^2 + e^{2P} Q,_\theta^2 \right) +  B^2 e^{(\lambda +\tau)/2}
\end{equation}
so that the AVTD solution is (since the equations for $P$ and $Q$ remain
the same)
\begin{equation}
\label{gowdyavtdsf}
P = v\,\tau, \quad \quad Q = Q_0, \quad \quad \lambda = -v^2 \,\tau
-v_\phi^2 \,\tau
\end{equation}
where $v_\phi = \pi_\phi/(2 \pi_\lambda)$. The magnetic potential will
now grow if $0 <
v^2 + v_\phi^2 < 1$. The additional scalar field kinetic energy means
that we can
have $0 < v < 1$ needed to keep $V_1$ and $V_2$ small with $v^2 +
v_\phi^2 > 1$ as is
needed to keep $V_3$ small. Thus it is expected that, eventually,
oscillations
will cease as $v$ and $v_\phi$ enter the required ranges at almost every
spatial point.

The results of numerical simulations of the magnetic Gowdy models (on
$T^3$ rather than
the solv-twisted torus \cite{weaver98}) with and without a scalar field
are shown in
Figures 4 and 5. In the former, the evolution toward the singularity of
$P$ at three
typical spatial points is shown. For the scalar field model, $v > 0$ at
all the spatial
points so that no further oscillations are expected. Once again, the
scalar field is seen
to suppress the bounces and the growth of small-scale spatial structure.
The entire
evolution $P(\theta,\tau)$ is shown in Fig.~5. The larger the initial
amplitude of
$\pi_\phi$, the more quickly the bounces will be suppressed. 

As in the spatially homogeneous case, the inhomogeneous models should
continue to
oscillate for scalar field potentials with an exponential form as
in(\ref{vscalarexp})
almost everywhere. Although the mechanism described here has been
examined only in
particular systems using arguments based on the MCP approximation
(reinforced by
numerical simulations of the full equations), we expect the behavior
discussed here to be
valid generically. The bottom line is that the scalar field kinetic
energy relaxes the
restrictions imposed by the Hamiltonian constraint on the gravitational
kinetic energy.
Since it is these restrictions which lead to the Mixmaster oscillations,
relaxing them
will allow an AVTD approach to the singularity.

\section*{Acknowledgments}
I would like to thank the Institute for Theoretical Physics at the
University of
California / Santa Barbara and the Institute for Geophysics and
Planetary Physics of
Lawrence Livermore National Laboratory for hospitality. I would also
like to thank A.
Rendall and A. Peet for bringing Refs.~\cite{belinskii73} and
\cite{lu96} respectively to
my attention and D. Garfinkle for providing a vacuum version of the code
used to simulate
the magnetic Gowdy models. This work was supported in part by National
Science Foundation
Grants PHY9800103 and PHY9407194. Some of the computations discussed
here were performed
at the National Center for Supercomputing Applications at the University
of Illinois.

\vfill
\eject

\section*{Figure Captions}
\bigskip
Figure 1. Comparison of Bianchi IX MSS trajectories for models with
(dashed line) and
without (solid line) a scalar field. The Kasner circle is shown. Each
saved data point of
the scalar field model is shown with a dot. The code described in
\cite{berger96c} is used to evolve spatially homogeneous Bianchi IX
models. The initial
scalar field momentum is
$p_\phi = .001$. Both models start from the same initial data (a). At
(b), it becomes
clear that the scalar field model trajectory cannot reach the Kasner
circle. The sequence
of Kasner solutions (indicated by values of $v_\pm$) terminates at (c)
for the scalar
field model. The vacuum model continues indefinitely although only the
first few bounces
are shown. The trajectory first reaches (c) for $\log_{10} |\Omega | =
3.45$.

\bigskip

Figure 2. Evolution of $\varphi$ toward the singularity for $U(1)$
symmetric vacuum
(solid line) and scalar field (dashed line) models at representative
spatial points. The
scalar field is modeled by $p_\phi = 5$ independent of space and time.
The simulation
consists of $128^2$ spatial grid points and is similar to those
described in
\cite{berger98c}.

\bigskip

Figure 3. Evolution of $\varphi$ toward the singularity for $U(1)$
symmetric (a) vacuum
and (b) scalar field models. A diagonal line $u = v$ in the $u$-$v$
plane is shown for
the same simulation as in the previous figure with $0 \le u,v \le 2\pi$
and $2 \le \tau
\le 39$.

\bigskip

Figure 4.  Evolution of $P$ toward the singularity for a magnetic Gowdy
model (see
\cite{weaver98}) with (dashed line) and without (solid line) a scalar
field. The initial
scalar field momentum is $p_\phi = \cos \theta$. There are $1024$
spatial grid points in
this simulation.

\bigskip

Figure 5. $P$ over the entire $\theta$-$\tau$ plane (a) without and (b)
with a scalar
field for the simulation of Figure 4 where $0 \le \theta-\theta_0 \le
2\pi$ for
$\theta_0 = -\pi/5$ and $0 \le \tau \le 61$.

\begin{figure}[bth]
\begin{center}
\makebox[4in]{\psfig{file=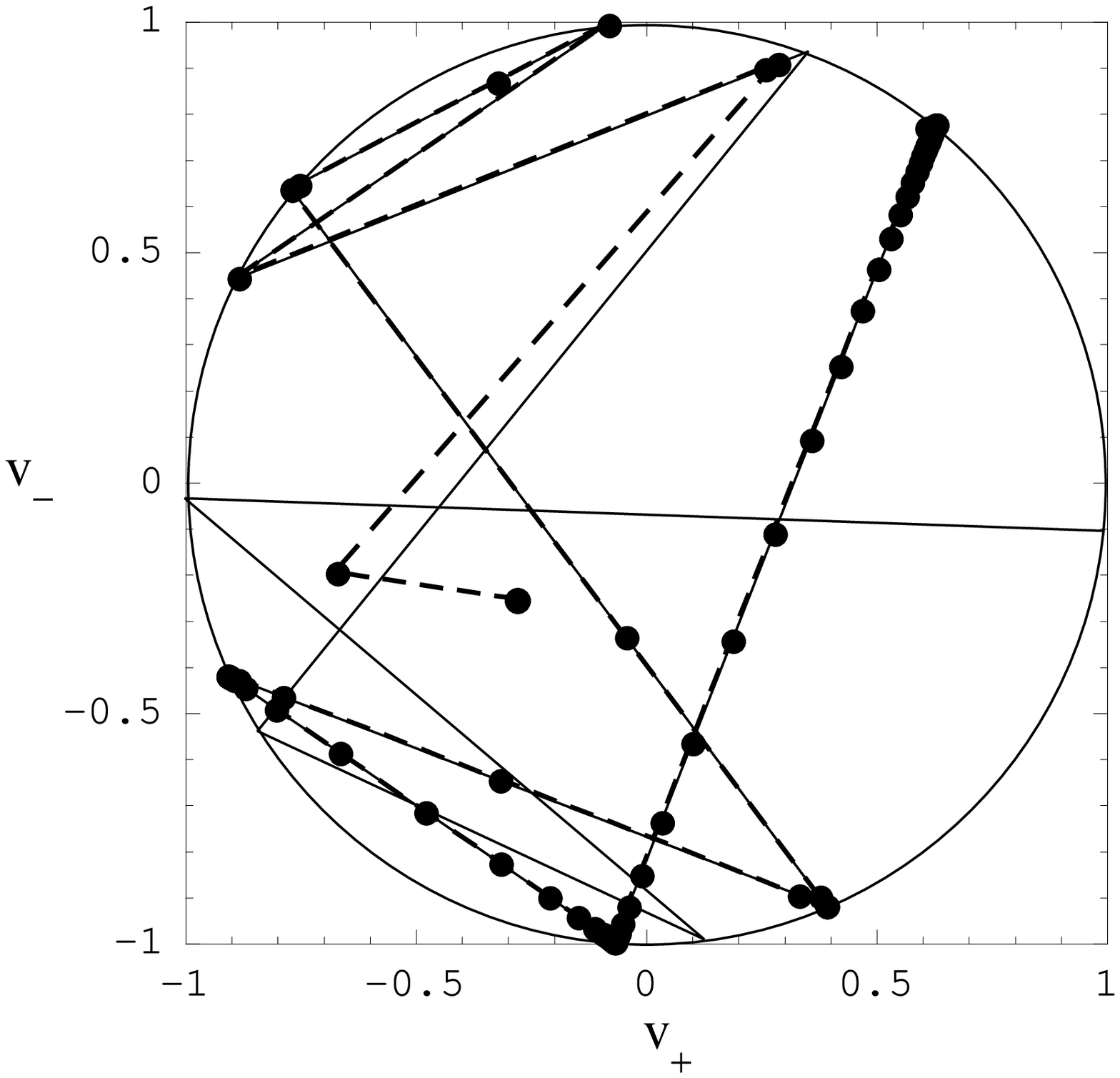,width=3.5in}}
\caption{}
\end{center}
\end{figure}
\begin{figure}[bth]
\begin{center}
\makebox[4in]{\psfig{file=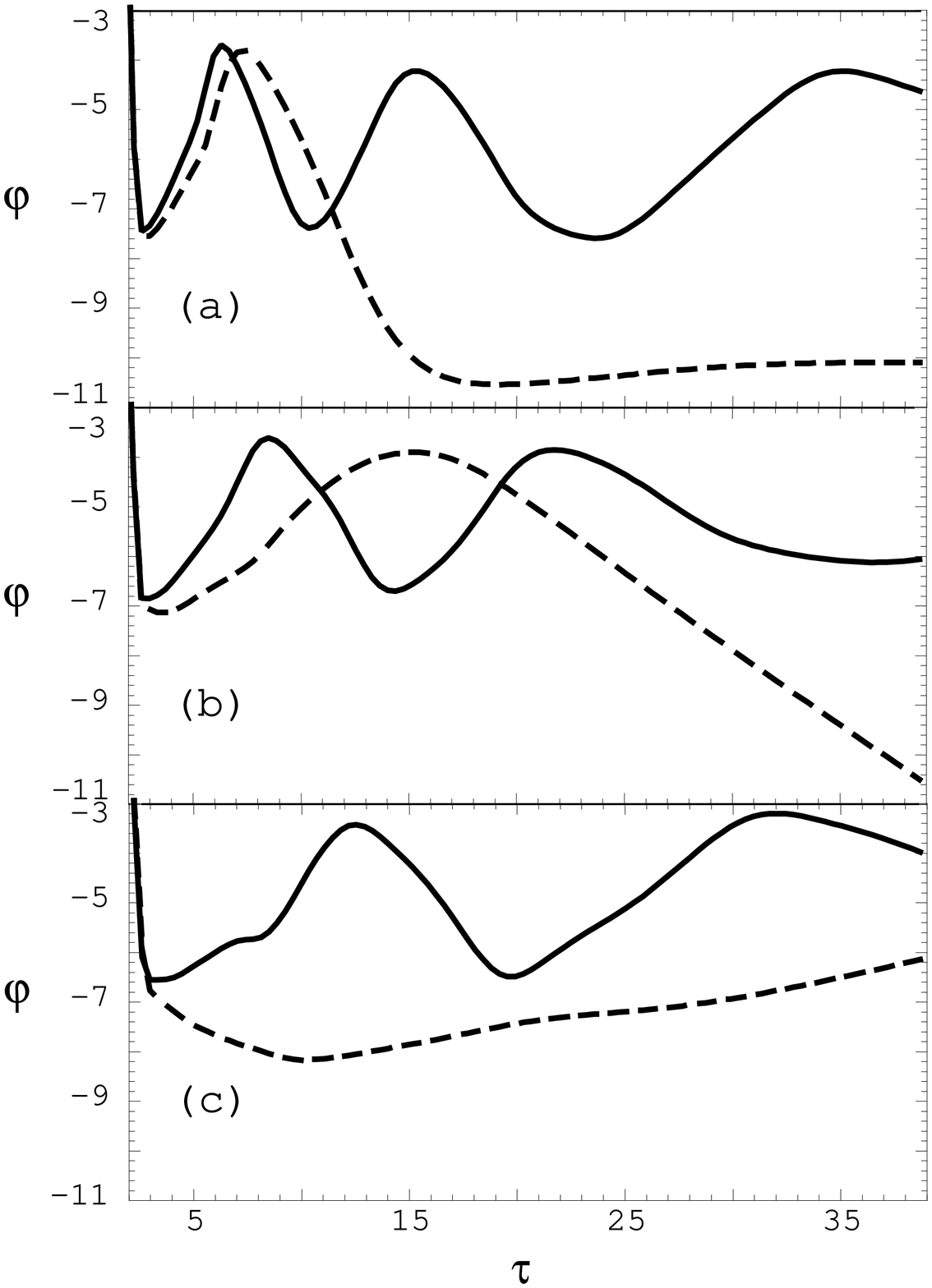,width=3.0in}}
\caption{}
\end{center}
\end{figure}
\begin{figure}[bth]
\begin{center}
\makebox[4in]{\psfig{file=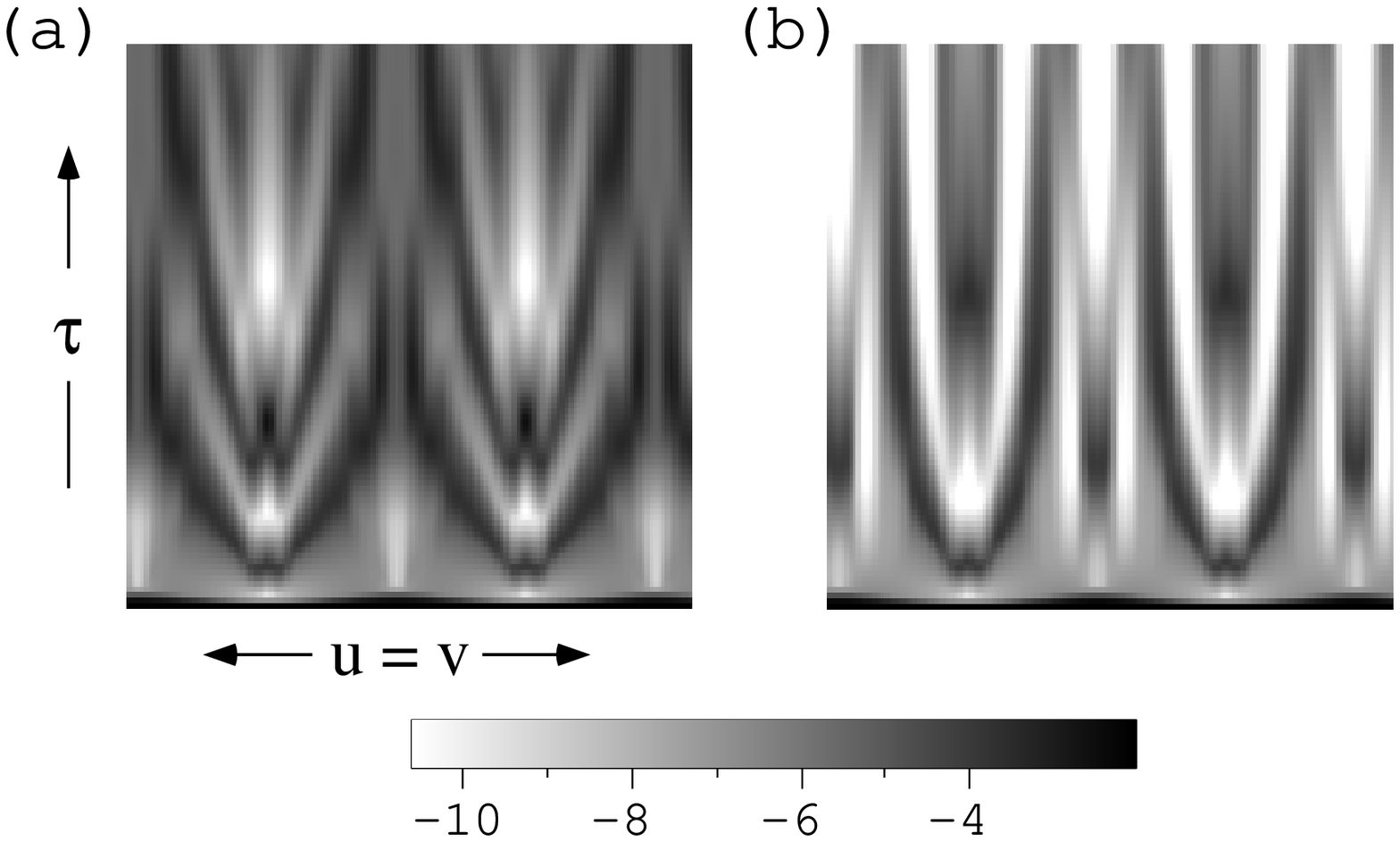,width=3.5in}}
\caption{}
\end{center}
\end{figure}
\begin{figure}[bth]
\begin{center}
\makebox[4in]{\psfig{file=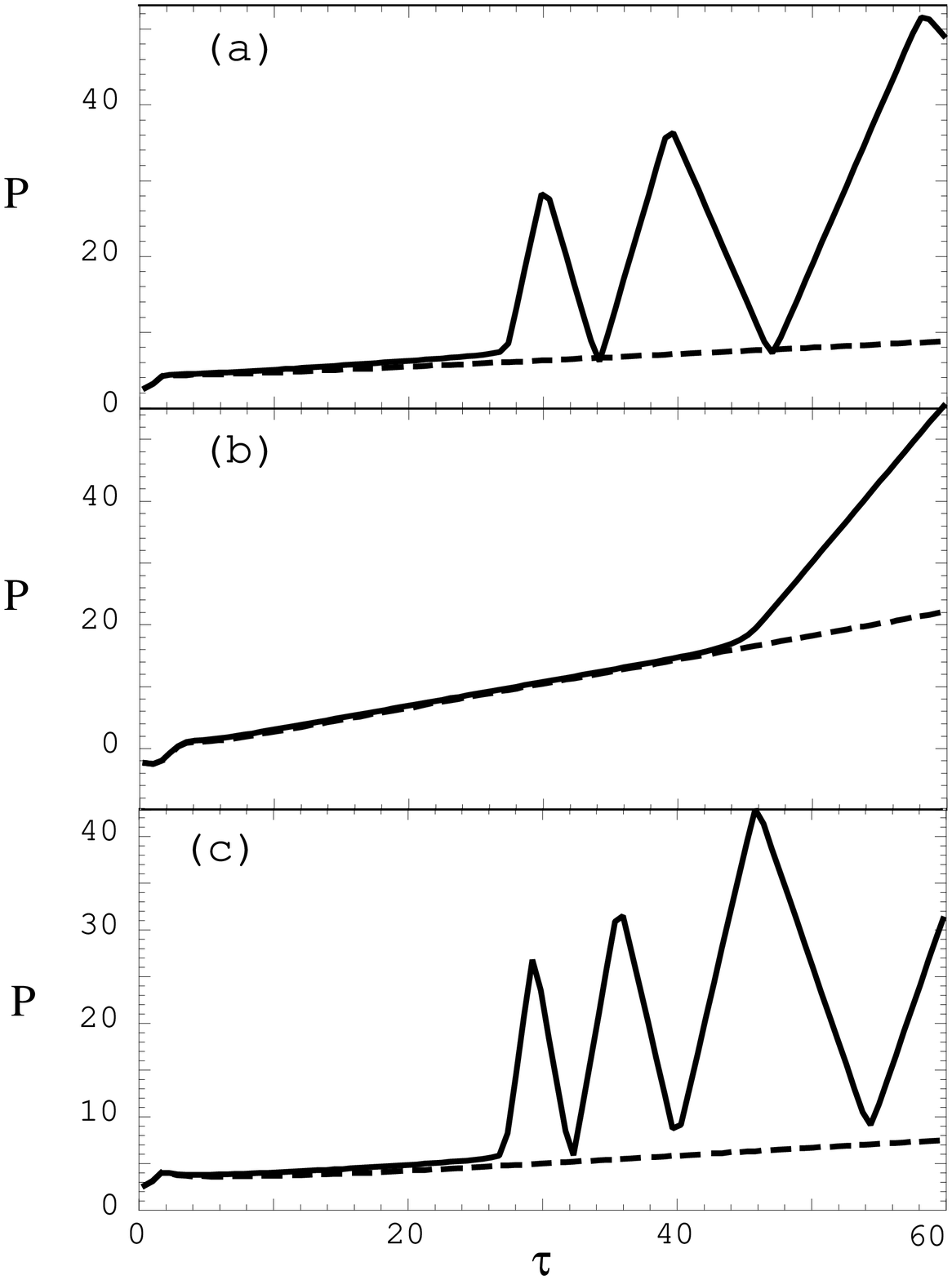,width=3.0in}}
\caption{}
\end{center}
\end{figure}
\begin{figure}[bth]
\begin{center}
\makebox[4in]{\psfig{file=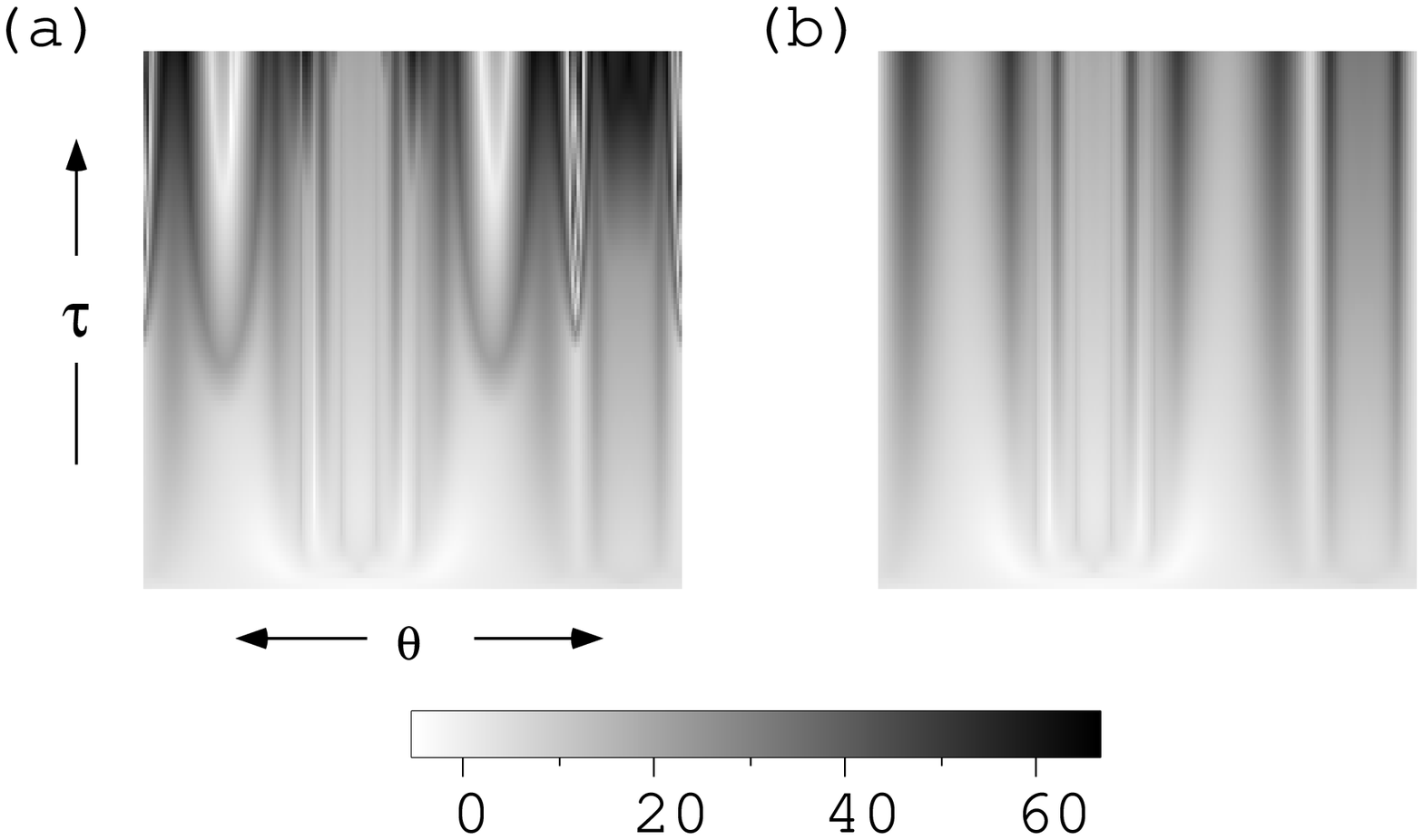,width=3.5in}}
\caption{}
\end{center}
\end{figure}


\bigskip

 
\begin{references}


\bibitem{belinskii69a}
{{Belinskii, V. A.} and {Khalatnikov, I. M.}},
{\em Sov. Phys. JETP},
{\bf 30},
{1174--1180} (1969).

\bibitem{belinskii69b}
{{Belinskii, V. A.} and {Khalatnikov, I. M.}}, 
{\em Sov. Phys. JETP}, 
{\bf 29}, 
{911--917} (1969)

\bibitem{belinskii71a}
{{Belinskii, V. A.} and {Khalatnikov, I. M.}},
{\em Sov. Phys. JETP},
{\bf 32},
 {169--172} (1971).

\bibitem{belinskii71b}
{Belinskii, V. A.}, {Lifshitz, E. M.}, and {Khalatnikov, I. M.}, {\em
Sov. Phys.
  Usp.} {\bf 13},  745  (1971).

\bibitem{belinskii82}
{{Belinskii, V. A.} and {Khalatnikov, I. M.} and {Lifshitz, E. M.}},
{\em Adv. Phys.},
{\bf 31},
{639--667} (1982).

\bibitem{misner69}
C.~W. Misner, {\it Phys. Rev. Lett.} {\bf 22},  1071  (1969).

\bibitem{kasner25}
Kasner, E., {\em
  Trans. Am. Math. Soc.}, {\bf 27}, 155--162,
  (1925). 

\bibitem{belinskii73}
{{Belinskii, V. A.} and {Khalatnikov, I. M.}},
{\em Sov. Phys. JETP},
{\bf 36},
{591--597} (1972).

\bibitem{grubisic93}
{Grubi\u{s}i\'{c}, B.} and {Moncrief, V.}, Phys. Rev. D {\bf 47},  2371
 (1993).

\bibitem{berger98d}
{{Berger, B. K.}, {Garfinkle, D.}, {Isenberg, J.}, {Moncrief, V.},
{Weaver, M.}},
{\em Mod. Phys. Lett.},
{\bf A13},
{1565--1574} (1998).

\bibitem{isenberg90}
{Isenberg, J. A.} and {Moncrief, V.}, Ann. Phys. 
(N.Y.) {\bf 199},  84  (1990).

\bibitem{berger96c}
{Berger, B. K.}, {Garfinkle, D.}, and {Strasser, E.},
{\em Class. Quantum Grav.}, {\bf 14},
L29--L36, (1997). 

\bibitem{lu96}
{{L\o{u}, H.} and {Pope, C. N.}},
{\em Nucl.Phys.},
{\bf B465},
{127--156}  (1996).

\bibitem{moncrief86}
Moncrief, V., {\em Ann. Phys. (N.Y.)}, {\bf 167},
118, (1986). 

\bibitem{berger93}
{Berger, B. K.}, and {Moncrief, V.},
{\em Phys. Rev. D}, {\bf 48}, 4676, (1993).

\bibitem{berger98b}
{{Berger, B. K.} and {Moncrief, V.}},
{\em Phys. Rev. D},
{\bf 57},
7235--7240 (1998).
Typographical errors appear in the full (but not asymptotic)
VTD solutions given in Eq.~(12) of this reference. The correct
expressions are $z =
-v_z (\tau -\tau_{0z}) -(1/2) \ln [|\mu_z| (1 +
e^{-4v_z(\tau-\tau_{0z})})]$, $\varphi =
-v_\varphi (\tau -\tau_{0\varphi}) -(1/2) \ln [|\mu_\varphi| (1 +
e^{-4v_\varphi(\tau-\tau_{0\varphi})})]$, $p_x = -4 \mu_z v_z$, $r = -4
\mu_\varphi
v_\varphi$.

\bibitem{berger98c}
{{Berger, B. K.} and {Moncrief, V.}},
{\em Phys. Rev. D},
{\bf 58},
064023  (1998) .

\bibitem{weaver98}
{Weaver, M.}, {Isenberg, J.}, and {Berger, B. K.},
{\em Phys. Rev. Lett.}, {\bf 80}, 2980 (1998).

\bibitem{weaver99}
M.~Weaver, Ph.D. Thesis, U.~of Oregon, 1999.

\end{references}
\end{document}